*Sequence analysis*

# High-speed and accurate color-space short-read alignment with CUSHAW2

Yongchao Liu[1,*], Bernt Popp[2] and Bertil Schmidt[1,*]

[1] Institut für Informatik, Johannes Gutenberg Universität Mainz, 55099 Mainz, Germany

[2] Institute of Human Genetics, University of Erlangen-Nuremberg, 91054 Erlangen, Germany

**ABSTRACT**

**Summary:** We present an extension of CUSHAW2 for fast and accurate alignments of SOLiD color-space short-reads. Our extension introduces a double-seeding approach to improve mapping sensitivity, by combining maximal exact match seeds and variable-length seeds derived from local alignments. We have compared the performance of CUSHAW2 to SHRiMP2 and BFAST by aligning both simulated and real color-space mate-paired reads to the human genome. The results show that CUSHAW2 achieves comparable or better alignment quality compared to SHRiMP2 and BFAST at an order-of-magnitude faster speed and significantly smaller peak resident memory size.

**Availability**: CUSHAW2 and all simulated datasets are available at http://cushaw2.sourceforge.net.

**Contact:** liuy@uni-mainz.de; bertil.schmidt@uni-mainz.de

## 1 INTRODUCTION

Unlike other sequencing methods, the SOLiD technology outputs short-reads in color space (a representation of two-base encoding) rather than in base space. In color space, each color represents two adjacent bases and each base is interrogated twice. This double interrogation causes a true single nucleotide polymorphism to result in a two-color change, while a sequencing color-error results only in a single color change. In some sense, this feature is helpful in discriminating true polymorphisms from sequence errors.

To align color-space reads to a nucleotide-based reference, base-space aligners cannot be directly used without decoding colors to bases. Although the known primer base enables data conversion, this approach has a significant drawback: a single color error will cause continuous base errors. Hence, an appropriate solution is to align color-space reads to a color-space reference. Thus, existing color-space aligners often use the approach of encoding a nucleotide-based reference as a color sequence in order to find the potential short-read alignment hits on the reference. These aligners may use different approaches to producing the final base-space alignments. One approach is identifying a final color-space alignment and then converting the color sequence to bases under the guidance of the alignment using dynamic programing (Li and Durbin, 2009). An alternative approach is directly performing a color-aware dynamic-programming-based alignment by simultaneously aligning all four possible translations of a read (Homer *et al*. 2009 and David *et al*. 2011).

We present a major extension of CUSHAW2 (Liu and Schmidt, 2012) for color-space short-read alignment. Our aligner is able to produce high-quality alignments at high speed, compared to existing top-performing color-space aligners.

## 2 METHODS

To improve mapping sensitivity, we have introduced a double-seeding approach, which combines maximal exact match (MEM) seeds as well as variable-length (ungapped/gapped) seeds derived from local alignments. For a single read, our aligner generally works as follows. First, all MEM seeds are generated for both strands based on our full-text minute index. Secondly, all mapping regions on the reference are determined through their corresponding seeds. All seeds are subsequently ranked in terms of optimal local alignment scores between the read and their corresponding mapping regions. Thirdly, dynamic programing is employed to identify the optimal local alignment of the read from the highest-ranked seeds. If satisfying the local-alignment constraints, including minimal percentage identity (default=90%) and read base coverage (default=80%), the optimal local alignment will be reported as the final color-space alignment. Otherwise, we will attempt to align the reads using semiglobal alignment. As an optimal local alignment usually indicates the most similar region on the reference, our semiglobal alignment takes the optimal local alignment as a variable-length seed, and computes a new mapping region on the reference. If the optimal semiglobal alignment satisfies the global-alignment constraints, including minimal percentage identity (default=65%) and read base coverage (default=80%), it will be reported as the final color-space alignment. This double-seeding approach enables us to rescue some alignments with more continuous mismatches and longer gaps. In such cases, we might fail to get good-enough optimal local alignments, as the positive score for a match is usually smaller than the penalty charged for mismatches and indels. For paired-end/mate-paired alignment, we generally follow the seed-pairing heuristic described in Liu and Schmidt (2012).

After obtaining a color-space alignment, we convert the color sequence into a base sequence with the dynamic programming approach proposed by Li and Durbin (2009). The translated base sequence will be re-aligned to the nucleotide-based reference using either local or semiglobal alignment depending on how its parent alignment has been produced.

## 3 RESULTS

We have evaluated the performance of CUSHAW2 (v2.4) by aligning both simulated and real color-space mate-paired reads to the human genome (hg19). This performance is further compared to that of SHRiMP2 (v2.2.3) and BFAST (v0.7.0a). We have ex-cluded BWA (Li and Durbin, 2009) and Bowtie (Langmead *et al.*, 2009) from our evaluations, as they have been shown to have inferior alignment quality compared to SHRiMP2 and BFAST in David *et al.* (2011). All evaluations are conducted in a workstation with a dual hex-core Intel Xeon X5650 2.67GHz CPUs and 96 GB RAM, running Linux (Ubuntu 12.04 LTS).

---

[*]To whom correspondence should be addressed.

We have used the sensitivity measure (calculated by dividing the number of aligned reads by the total number of reads) for both simulated and real reads. In addition, three other measures: recall, precision and *F*-score have been used on simulated reads, as the true mapping positions are known beforehand. Recall (precision) is calculated by dividing the number of correctly aligned reads by the total number of reads (by the number of aligned reads), and *F*-score is defined as $2 \times recall \times precision / (recall + precision)$. For a simulated read, it is deemed as correctly aligned if the mapping position has a distance of $\leq 10$ to the true position. As all evaluated aligners support mapping quality scores, they have further been compared by only considering the alignments with mapping quality scores $\geq 30$ (Q30). The runtimes are measured in wall clock time by running each aligner with 12 threads. Detailed alignment parameters are given in the supplementary data.

We have simulated two mate-paired datasets (read lengths are 36 and 50 respectively) from the human genome using the ART (v1.0.1) simulator (Huang *et al.* 2012) to evaluate all aligners (see Table 1). The 36-bp (50-bp) dataset contains 4,357,168 (3,137,161) read pairs and each dataset has an insert-size of $200 \pm 20$. In terms of alignment quality, for the 36-bp dataset, CUSHAW2 is superior to all other aligners for all measures and for both with and without Q30. BFAST performs better than SHRiMP2 for each case. For the 50-bp dataset, CUSHAW2 yields the best *F*-score and SHRiMP2 gives the best precision for each case. As for sensitivity and recall, CUSHAW2 performs best with Q30, while BFAST is the best without Q30. In addition, it is observed that for the 50-bp dataset, BFAST has a sharp performance drop, with respect to sensitivity, recall and *F*-score measures, from the evaluation without Q30 to with Q30. This suggests that BFAST has little confidence in the correctness of most of the reported alignments. Furthermore, this reflects that most of the reported alignments have higher probabilities to be false positives.

In addition, two real mate-paired datasets (with accession numbers SRR042786 and SRR064364 in NCBI SRA, respectively) have been used to evaluate all aligners (see Table 1). Each dataset comprises 50-bp reads and has an insert-size of $1500 \pm 500$. As SHRiMP2 and BFAST are quite slow, we have only used the first 6,000,000 read pairs of each dataset in our evaluations. For each dataset, BFAST yields the best sensitivity without Q30 and CUSHAW2 is the second best, whereas the latter performs best with Q30 and the former is second. For each case, SHRiMP2 yields the worst sensitivity.

In terms of speed, CUSHAW2 runs between one and two orders of magnitude faster than SHRiMP2 and BFAST. On the simulated datasets, CUSHAW2 is $71.8\times$ ($50.5\times$) faster for the 36-bp (50-bp) dataset than SHRIMP2 and $19.1\times$ ($36.5\times$) faster than BFAST. On the real datasets, CUSHAW2 yields a speedup of 12.5 (13.0) for dataset SRR042786 (SRR064364) over SHRiMP2 and 12.8 (12.8) over BFAST. As for peak resident memory size, SHRiMP2 takes the most memory of 40.5 GB and BFAST is the second most of 31.2 GB. CUSHAW2 has the smallest memory size of 4.7 GB, which is $8.6\times$ less than SHRiMP2 and $6.6\times$ less than BFAST.

**Table 1.** Alignment quality, runtimes (in minutes), and peak resident memory sizes (in GB) of all evaluated color-space aligners

| Dataset | Measure | CUSHAW2 | SHRiMP2 | BFAST |
|---|---|---|---|---|
| Simulated 36-bp | Sensitivity | **69.09**/**64.39** | 17.24/17.24 | 21.79/18.42 |
| | Recall | **68.17**/**63.52** | 15.02/15.02 | 19.73/17.94 |
| | Precision | **98.66**/**98.65** | 87.14/87.14 | 90.52/97.40 |
| | *F*-score | **80.63**/**77.28** | 25.62/25.62 | 32.40/30.30 |
| | Time/Mem. | **4**/**4.4** | 305/40.4 | 81/31.2 |
| Simulated 50-bp | Sensitivity | 67.07/**63.71** | 57.50/57.50 | **84.15**/12.36 |
| | Recall | 66.16/**62.86** | 57.17/57.17 | **70.94**/11.72 |
| | Precision | 98.65/98.65 | **99.42**/**99.42** | 84.31/94.81 |
| | *F*-score | **79.21**/**76.79** | 72.60/72.60 | 77.05/20.86 |
| | Time/Mem. | **4**/**4.5** | 201/40.4 | 145/31.3 |
| SRR042786 | Sensitivity | 49.89/**48.44** | 46.98/46.98 | **87.32**/26.42 |
| | Time/Mem. | **22**/**4.7** | 271/40.5 | 277/31.2 |
| SRR064364 | Sensitivity | 54.70/**53.21** | 47.28/47.28 | **87.09**/30.18 |
| | Time/Mem. | **25**/**4.7** | 325/40.5 | 320/31.2 |

For each alignment quality measure, the value $x/y$ (in percentage) means that $x$ is calculated from all reported alignments and $y$ from the alignments with mapping quality scores $\geq 30$. All best values have been highlighted in bold.

## ACKNOWLEDGEMENTS

*Conflicts of interest*: none declared.

# Supplementary Data for CUSHAW2 Color-space Alignment

Yongchao Liu[1], Bernt Popp[2] and Bertil Schmidt[1]

[1]Institut für Informatik, Johannes Gutenberg Universität Mainz, 55099 Mainz, Germany

[2]Institute of Human Genetics, University of Erlangen-Nuremberg, 91054 Erlangen, Germany

## 1  Alignment Parameters

CUSHAW2 employs the default parameters to construct color-space genome indices. As the alignment quality of BFAST is sensitive to both the number of genome indices and the spaced-seed masks, we have constructed 10 genome indices using the 10 spaced-seed masks, as recommended in Homer *et al*. (2009), in order to yield high alignment quality. The following list the command lines for BFAST genome index construction. In addition, Table S1 gives the alignment parameters of all evaluated aligners on both simulated and real datasets.

- bfast index -A 1 -f hg19.fasta -m 1111111111111111111 -w 16 -i 1 -n 12
- bfast index -A 1 -f hg19.fasta -m 111110111011101010010101011011111 -w 16 -i 2 -n 12
- bfast index -A 1 -f hg19.fasta -m 10111101011010010110000110100011111111 -w 16 -i 3 -n 12
- bfast index -A 1 -f hg19.fasta -m 10111001101001100100111101010001011111 -w 16 -i 4 -n 12
- bfast index -A 1 -f hg19.fasta -m 1111101101110111101111111 -w 16 -i 5  -n 12
- bfast index -A 1 -f hg19.fasta -m 11111110010100100010111110111 0111 -w 16 -i 6 -n 12
- bfast index -A 1 -f hg19.fasta -m 111101011100101000101011010101 11111 -w 16 -i 7 -n 12
- bfast index -A 1 -f hg19.fasta -m 111101101011011001100000101101001011101 -w 16 -i 8 -n 12
- bfast index -A 1 -f hg19.fasta -m 1111011010001000110101100101100110100111 -w 16 -i 9 -n 12
- bfast index -A 1 -f bfast-hg19.fasta -m 11110100101101101011100101101 11011 -w 16 -i 10 -n 12

Table S1. Alignment parameters of all evaluated aligners

| Dataset | Aligner | Parameters |
|---|---|---|
| Simulated | CUSHAW2 | -t 12 -avg_ins 200 -ins_std 20 |
| | SHRiMP2 | --no-qv-check -N 12 --insert-size-dist 200,20 --max-alignments 1 -1 filename_F3.fq -2 filename_R3.fq |
| | BFAST | match     -n 12 -A 1<br>localalign     -n 12 -A 1<br>postprocess     -n 12 -A 1 -v 200 -s 20 -Y 1 |
| Real | CUSHAW2 | -t 12 -avg_ins 1500 -ins_std 500 |
| | SHRiMP2 | --no-qv-check -N 12 --insert-size-dist 1500,500 --max-alignments 1 -1 filename_F3.fq -2 filename_R3.fq |
| | BFAST | match     -n 12 -A 1<br>localalign     -n 12 -A 1<br>postprocess     -n 12 -A 1 -v 1500 -s 500 -Y 1 |